# AtOMICS: A neural network-based Automated Optomechanical Intelligent Coupling System for testing and characterization of silicon photonics chiplets

Jaime Gonzalo Flor Flores[1], Connor Nasseraddin[1], Jim Solomon[1], Talha Yerebakan[1], Andrey B. Matsko[2], and Chee Wei Wong[1]

[1] Fang Lu Mesoscopic Optics and Quantum Electronics Laboratory, University of California, Los Angeles, CA 90095, USA

[2] NASA, Jet Propulsion Laboratory (JPL), Pasadena, CA 91109, USA

E-mail: jflorflores@ucla.edu; cheewei.wong@ucla.edu

**Abstract:**

Recent advances in silicon photonics promise to revolutionize modern technology by improving performance of everyday devices in multiple fields [1]. However, as the industry moves into a mass fabrication phase, the problem of effective testing of integrated silicon photonics devices remains to be solved. A cost-efficient manner that reduces schedule risk needs to involve automated testing of multiple devices that share common characteristics such as input-output coupling mechanisms, but at the same time needs to be generalizable to multiple types of devices and scenarios. In this paper we present a neural network-based automated system designed for in-plane fiber-chip-fiber testing, characterization, and active alignment of silicon photonic devices that use process-design-kit library edge couplers. The presented approach combines state-of-the-art computer vision techniques with time-series analysis, in order to control a testing setup that can process multiple devices and can be easily tuned to incorporate additional hardware. The system can operate at vacuum or atmospheric pressures and maintains stability for fairly long time periods in excess of a month.

1.   Introduction

With the growth of silicon photonics commercialization and foundries, new challenges in packaging and testing should be addressed to fully integrate the technology. While the cost of packaging for traditional integrated circuits (ICs) is variable, the greatest cost reductions come from optimizations in the process which achieve high throughput [2]; this includes the selection of appropriated processes that can be automated. In the case of testing, prices vary from a flat rate of 2% of the total manufacturing cost up to an order-of-magnitude higher [3], depending on the amount of testing performed and the reductions in liability cost that it could provide. However, packing and testing of photonic devices is usually more expensive because optical components need to be aligned to couple the light source. There are three main options to couple an external light source to a silicon photonics chip. This could be by grating couplers, evanescent coupling, or edge coupling, which does not use a large die area, and usually provides the highest input power [4]. In this paper we present an "*AuTomated Opto-Mechanical Intelligent Coupling System*" (AtOMICS) which combines testing hardware control, state-of-the-art computer vision algorithms, and customized time-series analysis models in order to generate a setup suitable for testing and characterization of silicon photonics chiplets and that can also be used for active alignment during the packaging process of laser-detector coupled devices. By replacing actuator drivers or adding additional data acquisition (DAQ) devices, AtOMICS is a flexible program that is readily compatible with other hardware too.

## 2. AtOMICS Description

The components of AtOMICS are shown in Figure 1a. The system is designed to couple input and output optical fibers at the edges of a silicon photonics chiplet, with sub-micrometer alignment tolerances. Designed inside a Janis ST-500 vacuum chamber, the system can operate at ambient pressure or vacuum, enabling the test of both regular silicon photonics devices and low pressure/temperature optical transducers in dual operation. The test results presented in the current paper have been collected in ambient or vacuum at ≈ 30 mTorr, and tests have been done with pressures as low as ≈ 1 µTorr. Figure 1b shows a zoomed-in image of the vacuum chamber and the internal components of the setup. The center gold stage is the chiplet mount, which includes a temperature controller, and is located on top of a $x$-$y$ piezoelectric positioner and a spacer block. The right and the left towers are a stack of $x$-$y$-$z$ positioners that move the input and output optical fibers to couple them to the waveguides on the silicon photonics chiplets. In our case, the piezoelectric stages are a set of two Attocube ANPx101s on the $x$- and $y$-axis, and an Attocube ANPz101 on the $z$-axis. Other stages can be used by replacing the controller driver on the program, as the algorithms and calculations are hardware independent. The entire structure is mounted on a support plate on top of a goniometer with a travel range of 10°. The goniometer is capable of moving the entire setup and is used as a way to input precise gravitational forces and induce changes in acceleration when testing photonic optomechanical sensors [5]. The vacuum chamber is closed by a lid that has a clear visor, and a microscope lens is mounted on top, as shown in Figure 1a. The microscope has three single-axis motors that can move it along the $x$-, $y$-, and $z$-axes, allowing it to reach the entire chiplet and the optical fibers. These motors are controlled by an ESP-300 controller and are equipped with relative encoders. The encoders are further tracked by AtOMICS and correlated to the piezoelectric stages position.

Figure 2 shows the rest of the elements of the system and how data is transmitted across them. The yellow arrows show the flow of light and experimental data collected in the optical carrier. Control signals are marked in blue and are bidirectional when data is read from the hardware and used as feedback to the software controller; movement confirmation is not taken as a return control signal in this case. The light blue arrows represent a physical input that is sent to the testing device, which could be the application of an external force or a magnetic field. The input side of the vacuum chamber has a lensed optical fiber that is connected to the laser source, which provides the optical excitation for the photonics devices, and is fed into a motorized polarization controller. The output fiber is connected to an optical switch, which is driven by an Arduino and connects via universal asynchronous receiver/transmitter. The optical switch sends the output optical signal to an external power meter that monitors the changes inside the vacuum chamber and provides a feedback signal to trigger the coupling controller. When the system is coupled and the power is stable, the switch sends the output signal to an optical detector that is connected to the data acquisition hardware; some examples of hardware supported by AtOMICS for data collection are oscilloscopes, frequency counters, spectrum analyzers, and multipurpose data acquisition hardware. The collected data from these instruments are later analyzed and transmitted to another hardware such as a real-time target for further integration with other testing systems.

AtOMICS integrates this hardware with a machine learning algorithm base on faster R-CNN for object detection and state-of-the-art object detection algorithms in order to automatically couple the devices which can be used for active alignment or testing of silicon photonics devices.

## 3. Conclusion

In this work we present an Automated Optomechanical Intelligent Coupling System suitable for testing of silicon photonic chiplets that uses edge coupling through lensed optical fibers. The system uses a variation of the Faster R-CNN object detection architecture combined with post-processing algorithms in order to determine the position of the different elements and track them as it moves them across the testing setup for automated and optimized laser chip coupling. The system is capable of precision fiber-to-chip alignment automatically and maintaining the alignment for time spans in the order of months tested. This system can aid in the self-testing and screening of silicon photonics devices in production and in applications where active long-term alignment is required.

**Additional Information**

Corresponding and request for material should be addressed to jflorflores@ucla.edu or cheewei.wong@ucla.edu.

**Acknowledgements**

We acknowledge discussions with Wenting Wang, Noah Himed, Yongjun Huang, Li-Hsu Yang and Jiagui Wu. The authors acknowledge the edge coupler test samples from the Institute of Microelectronics, with Dim-Lee Kwong, Patrick Guo-Qiang Lo, and Mingbin Yu. The authors acknowledge support from NASA and JPL through the Small Spacecraft Technology Program (SSTP).

**Conflict of Interest**

The authors declare no conflict of interest.

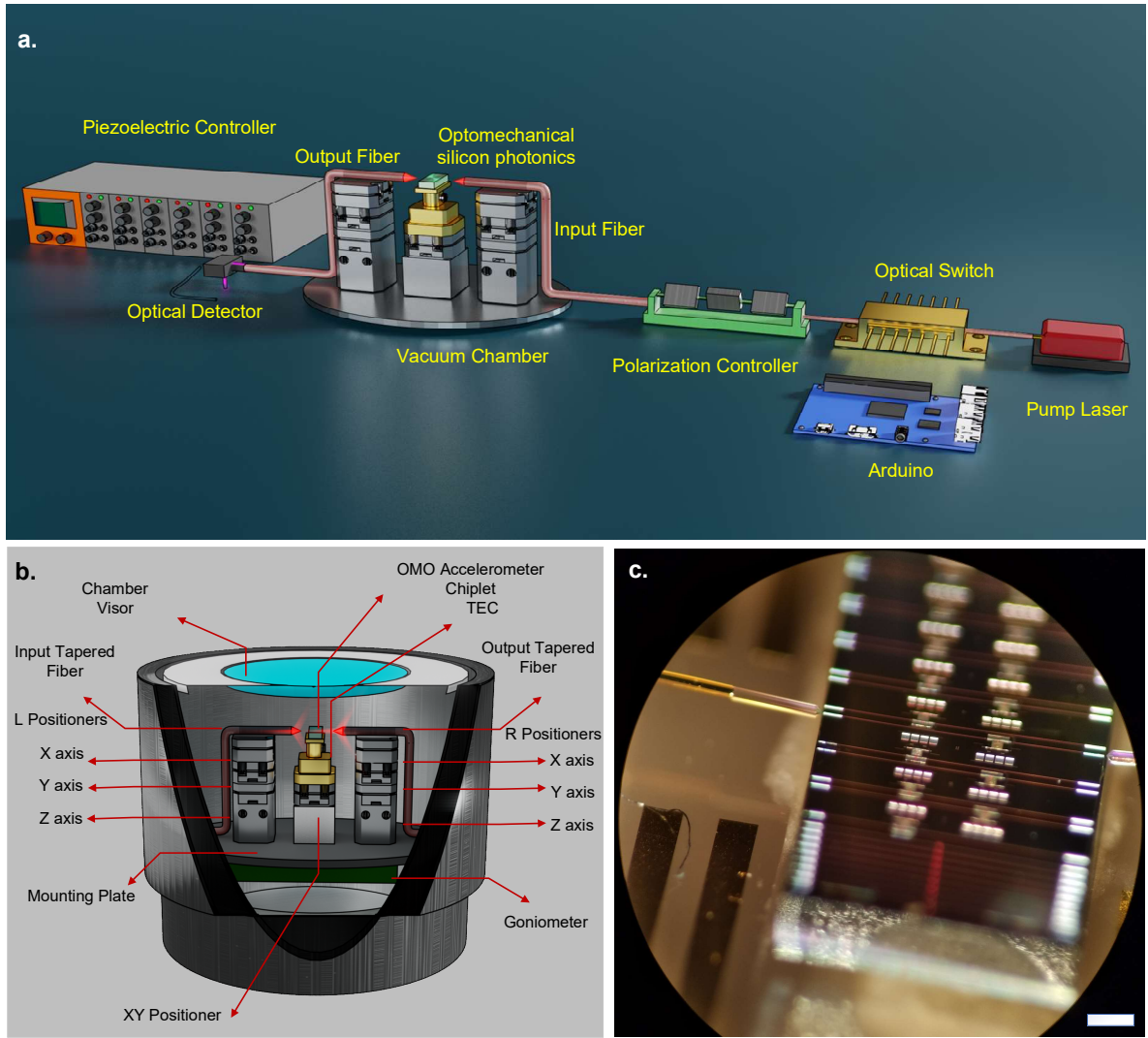

**Figure 1 | AtOMICS hardware setup and edge coupling. a,** Hardware integrated to AtOMICS. The chiplet is mounted inside the vacuum chamber and is coupled to a pair of lensed fibers. The microscope and CCD cameras monitor the positions of the optical fibers and the inverse tapered couplers. The system integrates a piezoelectric controller that translates the stages inside the chamber and a microscope controller that translates the camera axis. A pump laser and a polarization controller are connected to the input optical fiber, and an optical switch connects the output signal with the power meter and other data acquisition equipment. **b,** A zoomed-in depiction of the vacuum chamber interior. Each optical fiber is mounted on a 3-stage tower of piezoelectric stages that translate in the *x-y-z* axes. The interior setup is assembled on a mounting plate, located

on a goniometer to tilt the chiplet assembly. **c,** The microscope image of a silicon photonics optomechanical chiplet coupled using the AtOMICS setup. The chiplet is composed of multiple silicon optomechanical transducers arranged into two vertical lines. The lensed optical fiber on the left of the chiplet is aligned to the inverse tapered coupler (in white). The optical waveguide, which goes from the left of the coupler to the silicon photonics optomechanical transducer, is shown in lighter red across the chiplet. Scale bar is 200 µm.

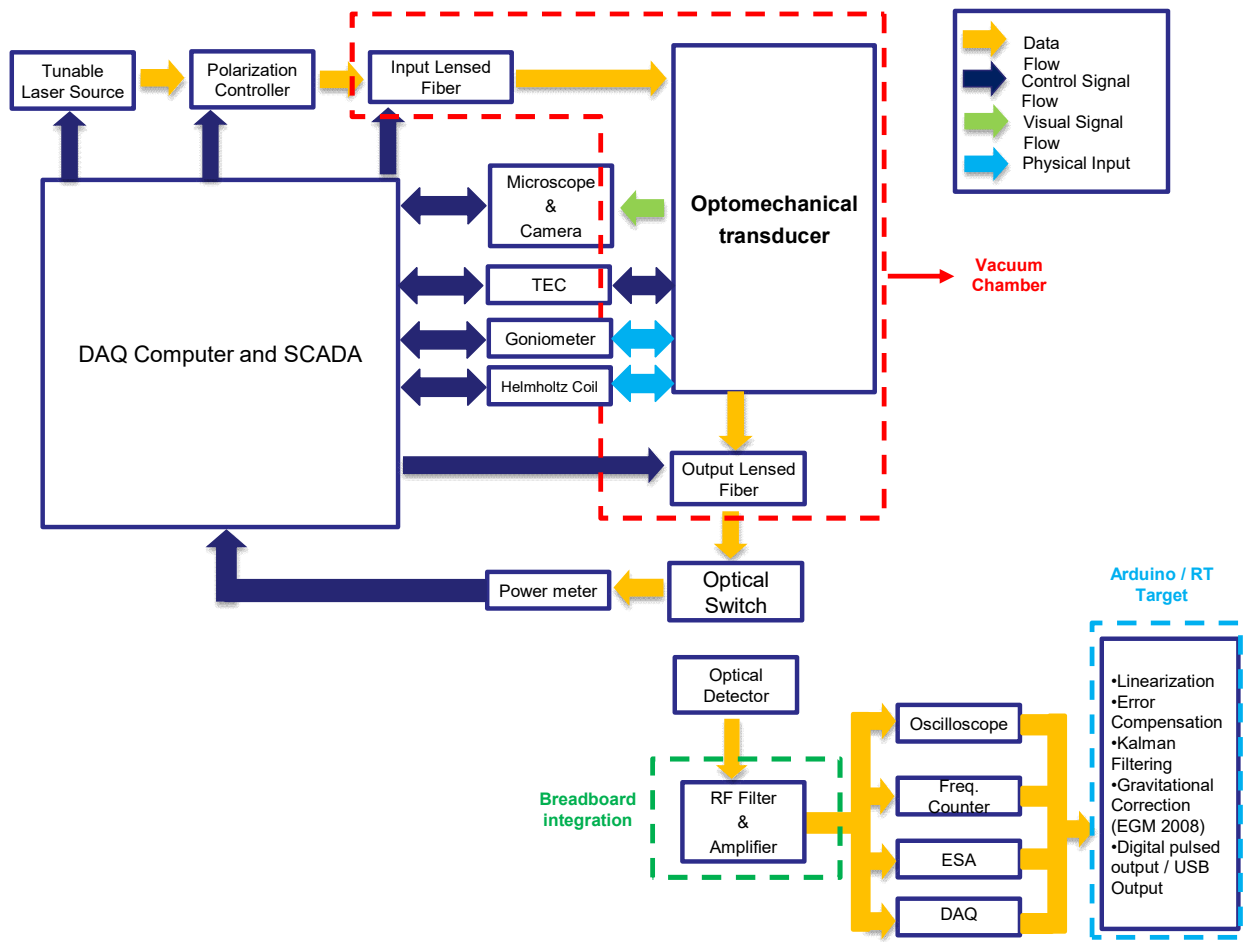

**Figure 2 | AtOMICS signal flow diagram.** AtOMICS is controlled by our in-house software that integrates multiple hardware components with control algorithms for coupling automation, supervision, and testing. Control signals are sent to the different components connected via universal serial bus (USB). Visual feedback is obtained from the CCD camera on top of the microscope, which is moved to different regions of the setup during the coupling and testing processes. The output signal is monitored by the power meter to provide feedback and optimized fiber position. Multiple data acquisition devices can be connected and managed by the system depending on the type of test that is being performed.